\documentclass[journal]{IEEEtran}
\usepackage{graphicx}
\usepackage{cite}

\usepackage{amssymb}
\usepackage{floatrow}
\usepackage{amsmath}
\usepackage{tikz}
\usetikzlibrary{arrows,shapes,positioning,shadows,trees}

\tikzset{
  basic/.style  = {draw, text width=2cm, drop shadow, font=\scriptsize, rectangle},
  root/.style   = {basic, rounded corners=2pt, thin, align=center,
                   fill=white!30},
  level 2/.style = {basic, rounded corners=4pt, thin,align=center, fill=white!60,
                   text width=5em},
  level 3/.style = {basic, thin, align=left, fill=white!60, text width=5em}
}

\ifCLASSINFOpdf
\else
\fi
\begin{document}
%
\title{Urban Sensing based on Mobile Phone Data: Approaches, Applications and Challenges}
%
%
%

\author{Mohammadhossein Ghahramani,~\IEEEmembership{Member,~IEEE,}
           MengChu~Zhou,~\IEEEmembership{Fellow,~IEEE} and Gang Wang
\thanks{This work is supported by FDCT (Fundo para o Desenvolvimento das Ciencias e da Tecnologia) under Grant 119/2014/A3.}
\thanks{}
\thanks{M. Ghahramani is a member of the Insight Centre for Data Analytics, University College Dublin, Ireland (e-mail: sphr.ghahramani@gmail.com, sepehr.ghahramani@ucd.ie).}
\thanks{G. Wang is with the Institute of Systems Engineering, Macau University of Science and Technology, Macau 999078, China and also China Telecom at Macau, Macau 999078, China, e-mail: (wanggang@chinatelecom.com.mo).}
\thanks{M. C. Zhou is with the Institute of Systems Engineering, Macau University of Science and Technology, Macau 999078, China, and also with the Helen and John C. Hartmann Department of Electrical and Computer Engineering, New Jersey Institute of Technology, Newark, NJ 07102, USA (e-mail: zhou@njit.edu).}

}

%
%

\markboth{}%
{Shell \MakeLowercase{\textit{et al.}}: Bare Demo of IEEEtran.cls for IEEE Journals}
%



\maketitle

\begin{abstract}
Data volume grows explosively with the proliferation of powerful smartphones and innovative mobile applications. The ability to accurately and extensively monitor and analyze these data is necessary. Much concern in mobile data analysis is related to human beings and their behaviours. Due to the potential value that lies behind these massive data, there have been different proposed approaches for understanding corresponding patterns. To that end, monitoring people's interactions, whether counting them at fixed locations or tracking them by generating origin-destination matrices is crucial. The former can be used to determine the utilization of assets like roads and city attractions. The latter is valuable when planning transport infrastructure. Such insights allow a government to predict the adoption of new roads, new public transport routes, modification of existing infrastructure, and detection of congestion zones, resulting in more efficient designs and improvement. Smartphone data exploration can help research in various fields, e.g., urban planning, transportation, health care, and business marketing. It can also help organizations in decision making, policy implementation, monitoring and evaluation at all levels. This work aims to review the methods and techniques that have been implemented to discover knowledge from mobile phone data. We classify these existing methods and present a taxonomy of the related work by discussing their pros and cons.
\end{abstract}


%
\IEEEpeerreviewmaketitle

\section{Introduction}\label{section.intro}
%
%
%
%
\IEEEPARstart{S}{martphones} are rapidly developing in recent years, and are becoming the central devices of communication and computing in people's daily life. This tremendous growth of usage has impacted the lives of people economically and socially for the better. Along with its development, mobile phone sensing has also achieved much popularity due to its convenience. These sensor-based devices can record conversations, movements, and activity states of individuals. Due to the widespread availability of smartphones and other mobile sensing-capable devices, sensor information has become very commonplace. Large data sets of human behavior are being collected and used to gain many insights into human interactions. They are utilized to target social activities, guide traffic, post advertisements, and support health care. For instance, they can be utilized in real-time monitoring of population density in urban areas or understanding the spread of diseases and accordingly provide procedural guidance. Furthermore, a smartphone has become a tool for economic growth and development. The extensive use of mobile applications has provided opportunities such as financial transactions through mobile devices (i.e., mobile payment), and entertainment applications. Reality Mining is a name coined for this data type exploration. It can be defined as a system's ability to regulate and extract a set of meaningful users' behavioral pattern \cite{steinbauer2012building}.

Smartphone data have been exploited in different directions, such as mobility path, city-wide sensing applications, traffic planning, and route prediction. Previous work on the utilization of these data accentuates the high potential of them in reading fine-grained variations of human's movements. However, there is a disconnection between high-level mobility path information and low-level location data. Hence, proposing an appropriate approach to deal with low-level location data and access meaningful users' mobility patterns is crucial. It is worth mentioning that there is a common assumption among all proposed methods in the literature: a definition of a mobility/interaction path to achieve cell phone users' mobility/interaction pattern at an abstraction level has been introduced \cite{bayir2009discovering}.

Smartphone data have distinctive characteristics that attract researchers and organizations to exploit them. The research undertaken in the past has resulted in different types of mobile sensing methodologies. They are based on position tracing or mobile positioning, i.e., tracing location coordinates of cell phones. Many Location-Based Services (LBSs) integrate Geographical Information Systems (GIS), Global Positioning Systems, and the Internet to suggest social activities and promotions. LBSs record people's movement, their flows, and events. Smartphone positioning can be categorized into active and passive approaches. The former is considered for handset tracking in which the device location is distinguished with a specific query by using radio waves (i.e., network-based methods such as Cell ID tracking and triangulation method) known as pinging. The latter analyzes data that are already stored via regular operations, i.e., billing data. This method needs the ability to carry out distance-based billing. The calls and SMSs sent or received generate records and containing cell IDs where they take place, allowing the phone's approximate location to be determined. By retrieving and analyzing such positioning information generated from mobile networks, mobile operators then gain significant insights for designing effective strategies. 

Various data/service provisioning approaches and applications have been also employed in mobile health, collaborative learning, and context-aware/location-based computing. We can categorize data/service combination into two distinctive directions, i.e., bottom-up and top-down. The former consists of an executable workflow, including known services, while the latter includes a non-executable scheme and a service selection phase. Given the highly heterogeneous characteristics of mobile computing, i.e., pervasive access to mobile services and ubiquitous communication among mobile devices, analyzing/tracing mobile data is not a trivial task. Consider the situation where cell phones are located outside of the communication range or when they are in offline mode. In these situations, mobile positioning and service provisioning are impossible. Therefore, an effective architecture for mobile service provisioning to address the challenges of service selection, e.g., avoiding frequent service recomposition, should be considered \cite{deng2017mobility}.

In this work, our goal is to review various techniques and methodologies that have been undertaken in the literature concerning smartphone data exploration. All the solutions that have been proposed by researchers to analyze people's behaviors and their consistent patterns are studied. We provide a typology of mobile phone data utilization in urban sensing domain, compare different analysis approaches and end-uses for decision-making systems. Providing a taxonomy of challenges and issues that require strict attention and careful considerations in the data acquisition and analysis phase is our concern. This work scrutinizes different proposed approaches/strategies and assesses existing challenges. We investigate their advantages and drawbacks and discuss various barriers that need to be dealt with.

The remainder of this paper is organized as follows: in Section \ref{section.approaches} different strategies for tracking and exploring mobile phone devices are classified. Section \ref{Urban.Dynamics} discusses the utilization of cell phone data for urban planning. Existing strategies and approaches for collecting and analyzing mobile phone data are introduced. Some case studies and empirical application of mobile phone data are provided in Section \ref{case.studies}. Section \ref{section.Challenges} presents potential challenges. Finally, Section \ref{section.conclusion} concludes this paper.




\section{Classification of Strategies}\label{section.approaches}
Most of the studies focusing on mobile phone data exploration aim to investigate human's positioning. Unlike other movement tracking techniques, such as road sensors, ticket tracking, and filling surveys, the collection of cell phone location data provides widespread coverage of the population in real-time. There are many methods for locating a mobile phone's position, e.g., using built-in components. The most renowned is satellite positioning using GPS. Other technologies such as Wi-Fi and Bluetooth can also be employed \cite{z.wang ,v.d.blondel, smit2012investigating}. Mobile phone positioning can be divided into two main categories: network-based positioning; and handset-based positioning. Given different characteristics of the two mentioned strategies, e.g., line-of-sight, and network coverage, the accuracy of a positioning technique can vary.

\subsection{Network-based Positioning}
This method includes cell activity and active/passive network querying. Inferring positions based on cell activity is a simple method to implement. However, because of uncertainty in spatial accuracy and the fact that this technique only counts handsets on a call, it is not a practical approach, and tracking populations can be biased. Therefore, to address this concern, active network querying methods (e.g., Round Trip Time, Angle of Arrival (AOA), and Triangulation) have been considered \cite{olsson2009sae}. Although the population's accurate locations can be polled in such methods, there are still some drawbacks, e.g., generating additional traffic to the network. Each phone should send information to a monitoring system, which could potentially increase the communication load on the network and the energy consumption of the handset. Since cellular networks are designed to deal with normal loads, there is probably inadequate capacity to handle the sequential pinging of all phones. Hence, it is impractical for tracking the entire population. Thus, it is only suitable for locating a small subset of handsets. Because of these kinds of problems, passive network querying techniques are needed.

When a phone is in its active mode (either calling or sending/receiving SMSs), its corresponding base station is logged continuously. In its idle mode, the information is stored once an hour. These data include the cell ID of the base station a handset is connected to and a time stamp. By passively scanning all of them, it is possible to track the locations of handsets in the network. This method is accurate to the nearest cell ID, can track journeys, and works wherever there is coverage. As mentioned above, the sample rate is around once per hour in idle mode, but can easily be increased by the carrier at the cost of additional network traffic. In other words, passive scanning can be used in conjunction with an active scanning method, in cases where there are handsets whose location information is needed more frequently. Table \ref{table:1} summarizes various network-based positioning methods in terms of their strengths and weakness. 

Since network-based strategies are applicable in the operator's side, most of the work in the literature have been focused on handset-based data sets and their relevant strategies. In the following sections, we study the methods that have been implemented based on hand-based data sets for exploring urban dynamics.

\begin{table*}[ht]
  \centering
  \footnotesize
    \begin{tabular}{  p{2.3cm} | p{4.6cm} | p{4.6cm} | p{4.6cm} }
    \hline
   {\bfseries Method}  & {\bfseries Description} & {\bfseries Strength}  & {\bfseries Weakness}\\ \hline
    Cell activity  & The simplest method for locating a mobile phone. &   Simple to implement;  No calculations are needed to obtain location information.& Not accurate enough in rural areas.  \\ \hline
    Angle of Arrival  & Network localisation technique based on angulation principle using an antenna array. & Localizing targets in a non-cooperative, and passive manner, which is highly desirable in sensor network applications.  It is a network-based method and supports legacy handsets.& Size and cost constraints of antennas; Requiring the line-of-sight.   \\ \hline
    Timing Advance & A method utilized to ensure that signals originating from a Mobile Station arrive at the Base Station at the correct time within the allocated time slot. & Does not consider processing times for each individual requests.& Using a coarse granularity measurement; It provides low range of resolution (550m). \\ \hline
    Received Signal Strength & A localization technique using signal strength. &  Simple to implement and low-cost method.& Lacking the accuracy. \\ \hline
    Time of Arrival (TOA) &  These measurements can be performed either at the base station or at the mobile station for position estimation. &  Supporting the legacy handsets, due to the network-based implementation. & Requiring synchronization between base stations and mobile stations; Suffering capacity problems due to the multilateral measurement principle.  \\ \hline
    Time Difference of Arrival (TDOA)& A triangulation technique that can be performed by both handsets and networks. & Requiring inexpensive and compact computing power; Accurate distributed timing synchronization. &  Lacking in locating narrowband and unmodulated signals. \\ \hline
    Observed Time of Difference Arrival (OTDA) & A TDOA-based approach designed to operate over wideband-code division multiple access (WCDMA) networks. & Solving the synchronization issues of base station's transmissions.& The accuracy of an individual time difference measurement depends on signal bandwidth. \\ \hline
    Enhanced Observed Time Difference & A TDOA-based location method based on the OTD. & Guaranteeing  the required synchronisation of the base stations; Providing high resolution.&  Requiring the handset's software modifications; Requiring additional mobile stations.\\ \hline
    Assisted global positioning system  & A method using both GPS and terrestrial cellular network localisation to obtain a geographic position. & High accuracy. & Requiring the line-of-sight. \\ \hline

  \end{tabular}
  \caption{Comparison of network-based techniques}
\label{table:1}
\end{table*}

\subsection{Handset-based Positioning}
Typically, handset-based data include handover records, Location data, and Call Detail Records (CDRs). Handover data are logs of a user's movement from a cell tower to another in an active call process. Location data include periodic location updates of cell towers. A mobile station controller (MSC) initiates a transition update in either the location register databases, i.e., home location register (HLR) or visitor location register (VLR), when a location variation is detected. Due to the lack of incentive for long-term storage, it is difficult to obtain HLR and VLR data from operators. In contrast, CDRs are easy to obtain as they are required for legal compliance \cite{bignami2007privacy} and thus stored for a long period. They contain information about all interactions between a mobile network and its subscribers that are needed for billing purposes. Among these data, there is also information on which base station subscribers are connected to. These data can be used to obtain valuable information about movements. 

Although mobile phone data is available at an operator's side, there are some difficulties for researchers to acquire them, most notably due to privacy concerns and business confidentiality issues \cite{gonzalez2008understanding}. As a result, some approaches have emerged which aim to address these issues by placing either embedded applications/sensors on a handset to log data, or by the construction of platforms in order to monitor data \cite{lane2010survey}. Among the prominent is the widely cited Reality Mining data set, an effort conducted at the MIT Media Laboratory. It follows near hundred subjects whose mobile phones are pre-installed with the applications that record and send data about call logs, Bluetooth devices in proximity of approximately five meters, cell tower IDs, application usage, and handset status. Subjects, including students and faculty, are observed by using these measurements over nine months. It also collects self-reported relational data from individuals \cite{eagle2006reality}. In \cite{shen2008mobivis}, the authors have utilized MIT data sets to present a visualization system for exploring the spatial and temporal data set. They have introduced a heterogeneous network to explore social-spatial data in a 2D graph visualization. A visual interface for performing semantic and temporal filtering is then proposed to support a large-scale cell phone data investigation. Ficek et al. \cite{ficek2010spatial} have proposed a method for locations data retrieval using the MIT dataset. They have conducted statistical analysis for such location measurements, i.e., people mobility patterns, spatial trajectories investigation and spatial-temporal data analysis. It should be mentioned that collecting data from embedded applications require the cooperation of handset owners to install applications to enable the logging procedures, which cannot be widely accepted, primarily owing to privacy concerns. 

\section{Urban Management}\label{Urban.Dynamics}
A better conception of when, where, and how individuals behave, particularly in populated regions, can lead to better urban infrastructure design. To that end, the dynamics of urban space and transportation should be explored. For example, understanding the flow of people and where they live is essential for urban planning. Such insights can help organizers to manage traffic flow and plan public transportation services. Innovative ways for assessing urban dynamics and human's behavior analysis with the use of mobile phone data have been considered. Smart cities incorporate pervasive and ubiquitous technologies to deal with environmental challenges. A multi-tier architecture for smart cities, consisting of various layers, e.g., human, service, infrastructure, and data layer, can be considered. All these layers should be interrelated. In this regard, relative efforts have been performed and different smart city perspectives, e.g., mobility and intelligent transportation, have been studied. In this section, the application of mobile phone data in achieving sustainable urban development is discussed. The aim is to explore whether and how research can support operations in cities by using a fine-grained data set. We intend to highlight various urban management functions, thorough application of smartphone data that have been employed to understand the increasing complexity of people settlements while considering the limitations and potentialities.

\subsection{Urban Dynamics}
The increasing penetration of mobile phones has made them attractive as urban monitoring sensors. When a mobile phone is handed over from one cell to another, an area in which the mobile phone is located can be traced. This capability/advantage of smartphones, e.g., spatial coverage, together with their high penetration in population can provide an opportunity to obtain valuable information cost-effectively. Both network-based and handset-based data sets can be utilized for analyzing urban dynamics. The former can help estimate the population within a cell's coverage area. And a pre-recorded database of signal strength fingerprints can be used to trace the mobility with handset-based data. The latter is more accurate but much time-consuming than the former.

Understanding an urban spatial structure has meaningful applications in a great variety of fields, including public transport and location-based recommendation. Thus, it is necessary to identify the relevant characteristics for a better understanding of such spatial structure. Research in this area aims to investigate dynamics by revealing the locations and intensities of urban activities and analyze spatial mobility patterns. In the field of urban spatial structures, it is required to analyze how human movement and activities impact an urban geographical space. Therefore, monitoring human movement is essential. In \cite{nikolopoulos2013reality}, the authors explain how data mining methods can be combined with large-scale multimedia storage. Their proposed approach can be helpful to mine large amounts of user-generated content (UGC) and gain insights into different perspectives of urban reality. They have presented three cases where UGC is employed to discover a citizen's perspective: city attractions, city issues/problems, and major events in the city. Chen et al. \cite{chen2014identifying} have proposed a popularity index of a channel to identify the hot-lines based on a CDR data set.  The density of users that travel across one channel and the diversity of travel behaviors are combined to infer each channel popularity level. In \cite{trasarti2015discovering}, the authors propose an analytical procedure intended to extract interconnections among different zones of a city, which emerge from highly correlated temporal variations of population local densities. First, a method to estimate the presence of people in different geographical areas is presented; then, they propose a method to extract spatial and temporal constrained patterns to obtain correlations among geographical areas in terms of considerable co-variations of the estimated presence. They have combined these two methods to deal with realistic scenarios of different spatial scale. Some work have proposed a set of models for inferring the number of vehicles moving from one cell to another using anonymous data \cite{becker2011tale, caceres2012traffic}. These models contain the terms related to a user's calling behavior and other characteristics of the phenomenon such as hourly intensity in cells and vehicles. A set of inter-cell boundaries with different traffic background and features have been selected for the field test.

Regardless of the benefits of these approaches, due to inherent characteristics of the mobile network geolocation, two consecutive spatial points to be measured might be separated by long distances and long periods. Then, corresponding trajectories may not be reliable and cannot be considered as a precise representation of individuals' real paths. To overcome these concerns, Calabrese et al. describe a real-time urban monitoring system that uses the Localizing and Handling Network Event Systems platform. This system is developed for the real-time evaluation of urban dynamics based on the anonymous monitoring of mobile cellular networks \cite{calabrese2011real}. Through the use of several probes, it extracts all the traveling signals and stores the measurements made by all active mobile phones. They have focused on visualization to monitor urban dynamics and to develop a real-time control system for cities.

\subsection{Understanding Mobility Flows}
Mobile phone data allow visualizing the flow of people throughout the entire urban system. They can be used to develop predictive models in a city-scale as a low-cost estimation for traffic. These data sets can help one perform urban management, route planning, traffic estimation, emergency detection, and general traffic monitoring. Moreover, mobile data can be regarded as operational information on cities' administration by aggregating people traces and collecting mobile phone traffic as a result of their behaviors. To capture mobility flows, some researchers \cite{demissie2013exploring, demissie2013intelligent, becker2011route} have used handover data collected from cellular towers. After pre-processing the data, they have studied flows through visualization software (e.g., GIS) and statistical analysis (e.g.,  classification algorithms). A qualitative interpretation of how the handover data can be useful in highlighting the flow of people in urban infrastructures have been provided via visualizations. It has been demonstrated that a high presence of people and cell towers with a high number of handovers are associated with each other. Moreover, the greater cell towers' proximity characterized by a high number of handovers denotes the greater movement. Notwithstanding the presence of associations between handover and traffic volume, however, there is a main limitation associated with this analysis: handover data is limited to mobile phones that are actively making calls, and the duration of the associated calls must be long enough to traverse the boundaries of two cells. Thus, it is not possible to make a direct correspondence between handover and traffic counts. These data sets are also coarse in space because they record locations at the granularity of a cell tower. Hence, analysis can be biased by temporal or spatial variations. In \cite{janecek2015cellular}, by utilizing the set of signaling events generated by active and idle devices, the authors have tried to overcome these drawbacks. While idle mobile phones provide a large volume of coarse-grained mobility data, active devices contribute with a fine-grained spatial accuracy for a limited subset of devices. The combined use of data from active and idle handsets enhances congestion detection efficiency in terms of accuracy, coverage, and timeliness. 

In \cite{zonghao2013resident}, the authors analyze different characteristics of human mobility by using billing data of more than one million anonymous users stored for seven days. They have proposed a method of recognizing the location of employment based on the regularity of individual trajectory. The residents' mobility is analyzed based on active cell phone data to observe partial mobility compared to overall mobility. Iqbal et al. have proposed an approach to implement OD matrices using traffic counts and CDRs \cite{iqbal2014development}. First, they analyze CDRs, including time-stamped cell tower locations and callers' IDs. Then, they use trips occurring within specified time windows to conduct tower-to-tower transient OD matrices for different periods. These matrices are associated with the corresponding nodes and transformed to node-to-node transient OD matrices. The actual OD matrices are estimated by using a microscopic traffic simulation platform. An optimization-based method is then implemented to specify the scaling factors that result in the best matches with the observed traffic counts. A methodology for passengers' demand estimation is presented in \cite{demissie2016inferring}. The significant ODs of inhabitants are extracted and utilized to build OD matrices. Thereafter, based on these routes the authors have claimed that strategic locations for public transport services can be reasonably suggested.  In \cite{toole2015path}, Toole et al. have presented algorithms to create routable road networks, generate verified OD matrices and trip summaries. They have routed these trips through road networks by using a paralleled Incremental Traffic Assignment algorithm. Aguilera et al. \cite{aguilera2014using} show that the specific conditions under which a cellular phone network is operated underground can make the passenger flows estimation possible in an underground transit system. They have conducted some experiments in an underground transit system to assess the potential of data for transportation studies with the help of a mobile network operator. They have also estimated the dynamic quantities improved, i.e., travel time, OD flows, and train occupancy levels from their cellular data set. The derived results are compared to those from Automatic Fare Collection data and direct field observations provided by the public transport authority.

Utilizing mobile phone data to reveal insights, e.g., OD matrices, is much faster than traditional surveying methods. However, there are serious concerns regarding employing them. 

\begin{itemize}
   \item Origin-destination matrices are representative of the devices connecting to the network at a given time. Consider a situation where a single cell tower covers a large area. In such circumstances, the intra-area movement cannot be traced. Hence, low sampling and penetration rates can negatively affect the validity of OD estimation. The integration of additional mobility information ideally can be considered to validate the revealed pattern.  

   \item Identifying the location where mobile phone owners live and work can be beneficial to infer their trips, behaviors, and consequently improve the validity of the analysis; however, there are privacy concerns. 
   
    \item We have observed that different hypotheses, e.g., uniform distribution and duration threshold, have been considered to identify activities. Such assumptions can bias the results since parameter-based models are highly sensitive to them. 

    \item Mobile network coverage depends on traffic and local topography. Defining the boundaries of the coverage area and taking the impact of them on constructing OD matrices into account are not trivial tasks.

    \item Handset-based data are generated when a subscriber is active, i.e., making or answering a call. Thus, the location of subscribers might not be updated and the analysis results seem to be biased by frequent users.

    \item Uneven distribution of mobile phones in a geographical region can negatively affect the analysis results.
   
\end{itemize}

Application of mobile phone data seems promising for exploring human mobility pattern, but more studies should be undertaken to validate the pattern and insights obtained from cell phone data in comparison with other approaches.
 
\subsection{Intelligent Transportation}
The data collected with travel questionnaires have been used to provide primary information for public transport providers, traffic planners, and infrastructure authorities \cite{demissie2013intelligent}. These data are the basis for routing, transportation modeling, and optimization. Acquired data can be regarded over a specific period and gives required information about travel behavior in different areas. A traffic information system (TIS) has two monitoring forms: sensor-based and cellular network monitoring. The former is expensive to deploy and maintain. It covers a small fraction of roadways. The latter can solve the issues of high cost and limited coverage but lacks accuracy. Traffic sensors, e.g., inductive loop detectors, magnetic sensors, video cameras, microwave radars, and infrared sensors, can be embedded in the pavement and collect data from all vehicles as they pass over them \cite{ban2010performance}. These fixed devices can count the number of people and vehicles passing a given point. They allow an operator to see and measure how traffic is flowing at a particular location. Their performance can be degraded by pavement deterioration, improper installation, and weather-related effects. The main drawbacks of these technologies include their cost (i.e., installation, maintenance, operation, and repair cost) and their restricted spatial coverage. To gain a realistic and complete view of traffic conditions, they must be installed in a large quantity. Therefore, they cannot be deployed globally at an acceptable resolution. Radio-frequency identification (RFID) transponders, GPS receivers, and mobile phones represent a novel way to monitor traffic data provided by vehicles.

Recently, intelligent transportation systems take a vital role. Its use can reduce traffic congestion and pollution. An intelligent traffic information system (ITIS) can provide individuals with valuable traffic data to support their route decision making. It takes advantage of the rapid advances in computers, sensors, and communication technologies. Driven by the fact that individual drivers are potential users of a mobile network, therefore, it is natural to consider them as the source of road traffic information. As a mobile network knows the approximate locations of active handsets, its data has the potential to revolutionize the study of city dynamics. Thus, the use of cellular data for intelligent monitoring of traffic has become popular. Understanding the mobility could take measures to better traffic management and provide governments with convenience to forecast the traffic demand. It could also lead to more precise decision making in the city and transportation planning process. There have been several studies on the use of mobile data to monitor road traffic with intelligent approaches. A typical mobile phone comprises several built-in micro-electro-mechanical systems (MEMS) sensors, e.g., accelerometer, magnetometer, GPS, and approximate network positioning that can be used for human mobility classification. GPS-based approaches have been commonly used to collect mobility-related information within a mobile network \cite{ahas2010daily,herrera2010evaluation}. GPS-equipped devices can compute the positions and instantaneous velocity readings of vehicles with high accuracy. They can either transmit their location data in real-time or store them in memory for later retrieval. In \cite{nitsche2014supporting}, by utilizing data obtained from smartphones, the authors present an approach to supporting travel surveys. They have classified the extracted features from the motion trajectory recorded by the positioning system and signals of an embedded accelerometer. Although the accuracy level of using these methods is high, their main drawback is the low penetration of the mentioned technologies in the population. Furthermore, vehicles equipped with a GPS device represent an added cost. Using it requires each phone to send information to a monitoring system, which could potentially increase the communication load and increases the energy consumption of the handset. Finally, it requires line-of-sight access to satellites, hence, unable to determine the accurate location while it is indoor.

The majority of literature in traffic monitoring via cellular networks targets non-real-time applications, such as the extraction of traffic flow statistics and origin-destination matrices for urban movement. Only a few studies \cite{calabrese2011real,ban2010performance,janecek2015cellular} address the specific problem of real-time road traffic estimation from cellular network signaling. Google Traffic is added as a feature on Google Maps to display traffic conditions in real-time on major roads and highways. But it works by analyzing the transmitted GPS-determined locations. As discussed earlier, there are some drawbacks regarding applying GPS-enabled technologies. It seems that an integrated system, one with consolidated phases comprising different layers such as traffic controllers, mobile communication systems, and the in-vehicle terminal, can ameliorate monitoring efficiently. By implementing an effective real-time monitoring system the information required to alert drivers to problems can be provided. Surveillance over a road, incident detection, and classification of vehicles are supplemental features that enable authorities to implement an efficient and convenient transport system which can detect threats and respond to security incidents to minimize risk.

Table \ref{table:2} summarizes the techniques and methodologies that have been utilized and reviewed in this section regarding various movement tracing methods, traffic sensing, and urban planning and compares them in terms of their pros and cons.

\begin{table*}[ht]
  \centering
  \footnotesize
    \resizebox{\linewidth}{!}{
    \begin{tabular}{  p{2.2cm} | p{6.0cm} | p{4.0cm} | p{4.0cm}  }
    \hline
   {\bfseries Work}  & {\bfseries Methodology} & {\bfseries Pros}  & {\bfseries Cons}\\ \hline
    \cite{ahas2010daily}, \cite{herrera2010evaluation}, \cite{shen2008mobivis} & GPS-equipped handset as probes to gather mobility related information within a cellular network. &  High Accuracy & Low penetration of GPS-equipped devices in population.   \\ \hline
    \cite{calabrese2011real}, \cite{tranos2012smart}, \cite{tranos2013mobile} &  A real-time representation of city dynamics through handover or cellphone trajectories from registered users. & Using pervasive computing. & High complexity.   \\ \hline
    \cite{caceres2012traffic}, \cite{toole2015path}, \cite{demissie2013exploring}, \cite{demissie2013intelligent}, \cite{becker2011tale}, \cite{Jiang2017Activity} & Utilizing handover and CDRs to estimate traffic volume and human mobility. & CDRs are relatively easy for mobile phone operators to collect.& Limiting the number of observable devices to a small fraction of the whole population.   \\ \hline
    \cite{jiang2011understanding}, \cite{jiang2012clustering}, \cite{becker2011tale}, \cite{Thuillier2017Clustering} & Clustering data into representative groups according to their daily activities. & Help urban management by answering when, where, and how individuals interact with different places. & Need to compose social networks and human interactions. \\ \hline
    \cite{zonghao2013resident},\cite{wang2010transportation}, \cite{Pinelli2016Data} & Analyzing spatial-temporal characteristics of human mobility via billing data. & Able to recognize the location of employment based on the regularity of individual trajectory.& Lacking real-time analysis. \\ \hline
    \cite{kuusik2010ability}, \cite{Zhou2018Support}, \cite{Demissie2019Trip}, \cite{Zhong2017Characterizing} & Investigating urban activity destinations and human travel patterns to monitor the concentration of people. &  Able to quantify the long-term effect of events in the context of destination marketing. &  Tracing by applying passive mobile positioning data can be biased by frequent users. \\ \hline
    \cite{iqbal2014development},\cite{friedrich2010generating}, \cite{demissie2016inferring}, \cite{Caceres2007Deriving}& Developing Origin-Destination matrices, using CDRs. & Able to detect the congestion. & Lacking real-time estimation. \\ \hline
    \cite{chen2014identifying}, \cite{Jahangiri2015Applying}, \cite{Lv2015Road}  & Proposing popularity index that utilizes diversity and density index of channels to identify the hot lines by using CDRs. & Exploring the human flows in an urban area with a quantitative measurement of an urban spatial structure. & Suffering from different spatial accuracy of cells. \\ \hline
    \cite{janecek2015cellular} & Exploiting the set of signaling events generated by both idle and active devices. & Able to overcome the limitation of a small number of observable devices.&  Increasing the communication load of the network.\\ \hline
    \cite{kang2013exploring}, \cite{Thejaswini2015Novel} & Analyzing population concentration by using GPS devices in transportation systems. & Able to provide high accuracy.& Requiring each phone to send information to monitoring system; and line-of-sight dependency. \\ \hline

  \end{tabular}}
  \caption{Comparison of different approaches for the mobile phone data analysis}
\label{table:2}
\end{table*}

\section{Empirical Applications of Mobile Phone Data}\label{case.studies}
\begin{figure}
  \includegraphics[width=\linewidth]{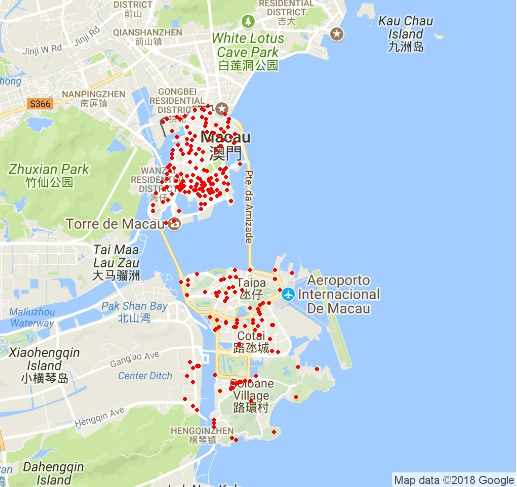}
  \caption{Distribution of cell towers in Macau (Source: Google Maps).}
  \label{fig:map1}
\end{figure}

Mobile phones are among the technologies that high-value solutions can be created from them. Significant changes in regular patterns of human manners could signal a quick response to an urgent situation, thus, monitoring behaviors could be taken into account to identify when and where an event has occurred. Given our discussion about positioning methods, we can divide mobile phone data into three main categories: 1) CDRs; 2) LBSs' data, and 3) handover data. These datasets contain the Spatio-temporal information of users. These features enable us to represent the intensity of different human behaviors through space and time. As illustrated in Fig. \ref{fig:map1}, we have located different cell towers based on a CDR obtained from a Telecommunication company in Macau. These spatial objects can be considered as points referenced by latitude and longitude and can be used to describe geographical patterns of interest. Different strategies regarding spatiotemporal clustering are discussed in more detail next.

\subsection{Spatial-Temporal Analysis}
Much of the worldwide data can be geo-referenced and consist of measurements or observations that are taken at specific locations, which indicates the importance of geospatial big data handling. Such data can be points referenced by latitude and longitude or within particular regions, so-called areal data. Their related studies aim to describe geographical patterns of interest. Positioning techniques can be used for obtaining the Spatio-temporal distribution of smartphones as the resolution of geo-location has been improved recently. These investigations have attracted significant attention, specifically in urban planning and transportation studies. Mobile phone interaction can be considered as a function of the overall population and observed spatial and temporal stationarity of different areas in a city. Given such data sets, we can identify mobile phone spatial and temporal pattern and its corresponding transformation based on population and density. By exploiting spatial and temporal data, i.e., coordinates of cell towers and their interactions, we can present a Spatio-temporal analysis model to capture the effect of urban density on transportation mode choices or evaluate trends of human behavior.

\begin{figure}
  \includegraphics[width=\linewidth]{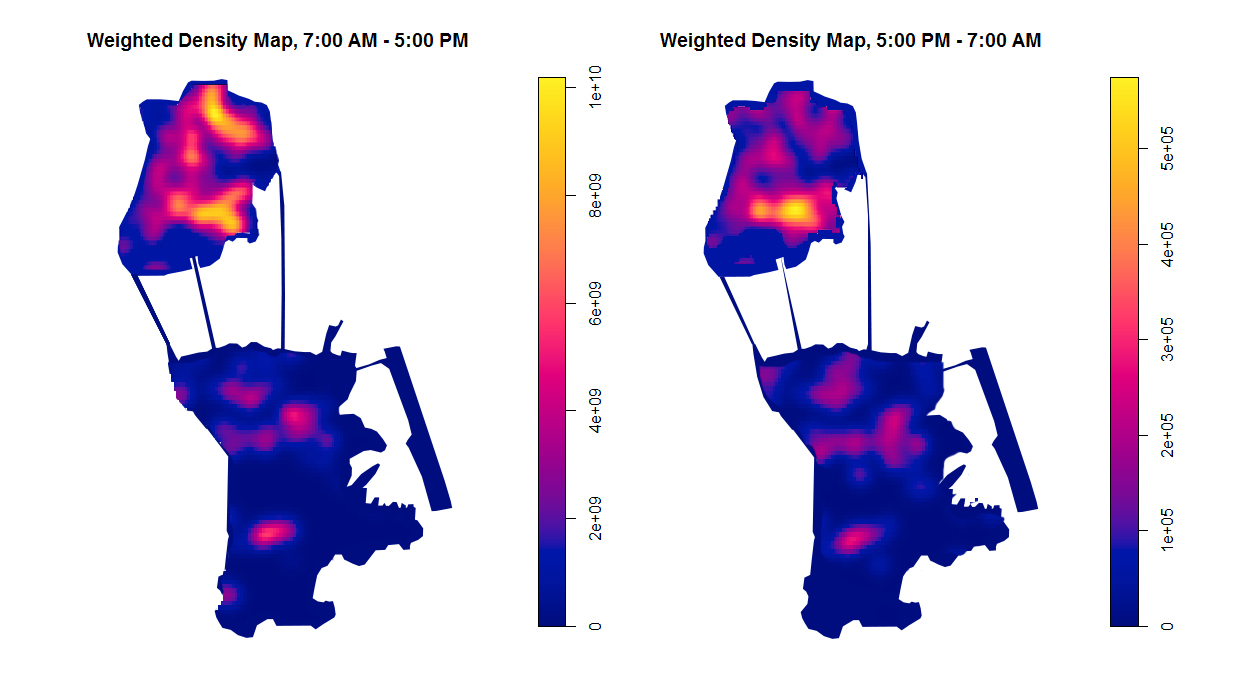}
  \caption{Spatial-Temporal analysis of mobile phone data in Macau.}
  \label{fig:hotspots}
\end{figure}

In \cite{ghahramani2018Hotspot}, we have utilized different correlation analyses to scrutiny the dynamics of a city. A descriptive spatial auto-correlation analysis (a global approach) is carried out to illustrate the relations among different areas. A local correlation measurement is then conducted to predict significant areas among cell towers. By determining spatial objects' clusters given the temporal characteristics of CDR, we have predicted the location of hotspots. A Kernel Density Estimation (KDE) method is then applied to the calling behavior dataset to depict these hotspots on the map. This mapping technique identifies the areas where there is a high level of activities in terms of calling patterns. Fig \ref{fig:hotspots} illustrates the results. We have considered the cell towers as the spatial objects and frequency of calls as variables.

\begin{figure}
  \includegraphics[width=\linewidth]{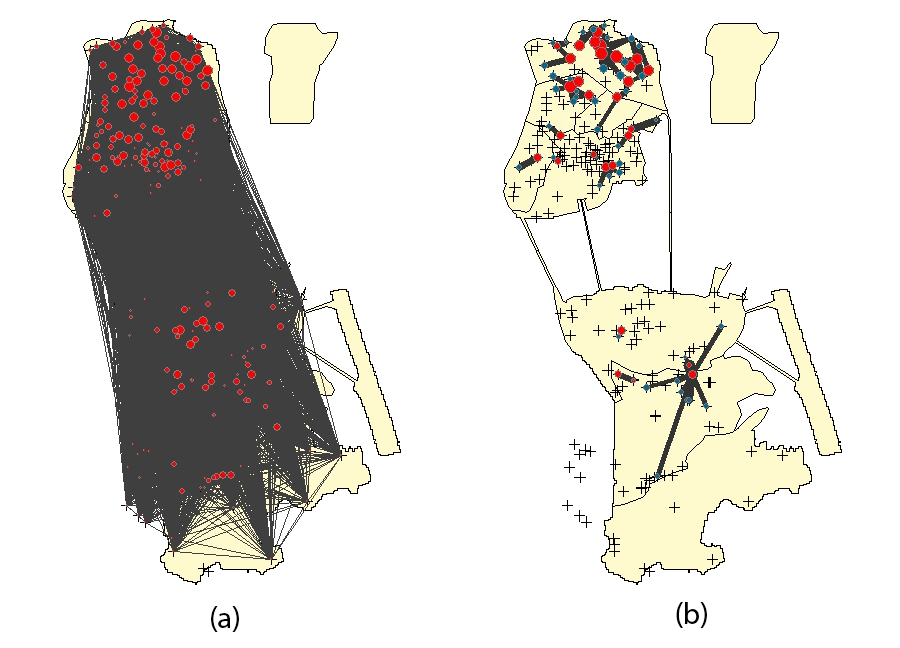}
  \caption{Mobile phone network analysis.}
  \label{fig:graph}
\end{figure}

The spatiotemporal analysis is more sophisticated than relational data processing in terms of algorithm efficiency and the complexity of possible patterns since an interrelated information at a spatial and temporal scale have to be considered. Mobile phone data can be used to interpret patterns embedded in the interaction flows of people. We can consider the geographical context of subscribers/cell towers to discover structures of interactions. Let's take the mobile phone interactions as a network graph with cell towers as its nodes and interactions as the edges. When coordinates of nodes are available, such networks can be considered as geographical networks, and the relationship among their components can be analyzed.
We can define $G = (V, E)$ be a call-network with $N$ nodes, where $V$ = \{$V_1, V_2, ..., V_n$\} is the set of vertices (cell towers), and $E \subseteq N \times N$, is the set of connecting edges. i.e.,

\begin{equation}
E=\begin{bmatrix}
         E_{ij}
        \end{bmatrix}_{N \times N}\end{equation}
where $i$ and $j$ represent cell towers $i$ and $j$. In line with this definition, we have implemented a Hierarchical Agglomerative Clustering (HAC) method on a CDR to detect interaction communities in \cite{ghahramani2018Extracting}. A HAC starts with each object (cell tower) in its cluster and then repeatedly merge similar clusters into broader ones. We have explored significant interaction patterns given the spatial heterogeneity of a mobile phone network. By implementing similarity measures, the proposed algorithm calculates the distance among clusters. These clusters are then merged until there is only one cluster remaining, or a certain termination condition is met. The spatial characteristics of nodes, together with an optimal level of the hierarchy is also proposed in our partitioning method. These insights can help organizations in decision-making and policy implementation. Fig. \ref{fig:graph}(a) illustrates the mobile phone network in Macau and interactions of cell towers. Fig. \ref{fig:graph}(b) reveals the community patterns detected through mobile phone interaction exploration.

\begin{figure}
  \includegraphics[width=\linewidth]{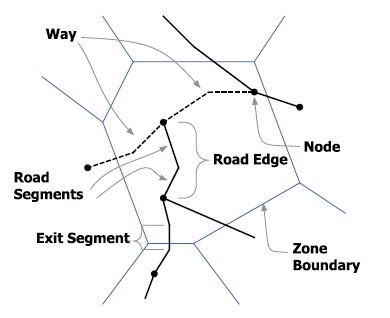}
  \caption{Segmentation of a road graph.}
  \label{fig:road}
\end{figure}

Dong et al. have analyzed social interactions by spatial modeling of the interplay between mobile phone subscribers' demographics and their social behavior \cite{dong2017user}. According to the results of the experiment demonstrated in their work, it is possible to predict users' gender and age by analyzing their calling behavior. By implementing a double-label classification model, they have shown how to infer subscribers' demographic information. They have defined two dependent variables, i.e. gender and age, and the correlation between those and other dependable features are modeled. In another work, Qiao et al. have implemented a spatiotemporal model based on a hidden-markov model to monitor the traffic \cite{Qiao2017real}. They have modeled urban road network as a graph. To that end, a junction intersecting roads are taken as the nodes while roads themselves regarded as the edges. Fig \ref{fig:road} reveals different road segments which are considered as the graph components. The Markov model is then adopted to infer hidden underlying structures of sequential traffic data on that road network. They have also defined the trip trajectories almost the same as the definition of the sequence $D_i$ presented in Section \ref{Urban.Dynamics}.

\section{Challenges}\label{section.Challenges}
We have presented how cell phone data can be utilized to gain intuitions into the complicated process of urban dynamics. We have outlined the mobile phone data applications with a particular focus on human movement, traffic sensing, and urban planning. The strengths and weaknesses of various approaches are given in each specific subsection and consequently are summarized in two tables to provide recommendations on different methods for different applications. Besides reviewing existing processing methods, their advantages and drawbacks are fully discussed. Some other generic challenges are summarized next.\\
\indent \textbf {Data access:} accessibility is probably the most remarkable hurdle to exploit mobile phone data because of the limited interest of governments and organizations to make them available as caused by privacy concerns. However, this can be changed by creating data standards that ensure data privacy. Providing network-based data can be costly to generate, and Telecom companies treat it as a commodity. Moreover, sharing mobile phone data sets can be a threat to private companies' business. Data deprivation can make sustainable development impossible.\\
\indent \textbf {Data quality:} the quality of data can be defined as the fitness of a data set for use in a specific domain. Take the Spatio-temporal analysis as an example. In such studies, fine-grained location data should be provided for applications such as location-based services, route planning, and transportation development. However, in rural regions, the spatial resolution may be poor. Data quality issues, e.g., lack of integrity constraints, inconsistent aggregating, would lead to reduced reliability and validity.\\
\indent \textbf {Privacy issue:} as discussed, the location awareness ability of mobile phones can make the geographical position of these devices available. Positions can be determined either independently through built-in components or externally by networks with which mobile phones connect to. Together with the benefits that this ability brings, there are myriad privacy implications. These logs can be stored and analyzed for multiple reasons (e.g., billing purposes, real-time routing assistance, destination guides, environmental condition, and wireless advertising) and might be disclosed. Such disclosure has non-technical and technical aspects \cite{Lin2018Frameworks, Zhang2016Privacy}. For example, traffic interactions can be intercepted by unauthorized parties. However, the sensitive information of people's communications must be preserved. People's mobility patterns can consist of private data that one does not want to be revealed. Hence, mobile phone data sets must be anonymized (i.e., using unique IDs or hashing techniques) when publicly available by removing names/numbers to preserve privacy.\\
\indent \textbf {Computing issue:} processing large amounts of mobile phone data may exceed the capacity of traditional analytic tools. Extracting meaningful insight from a massive data set can cause a processing issue. Traditional data architectures cannot handle a large volume of mobile phone data since they are not able to deal with different characteristics of massive data sets (e.g., velocity, variety, and veracity). This inability has led to the development of Big Data analytics platforms, and Cloud-based and Edge Computing  \cite{Casadei2019development, Liu2017Scalable} methodologies seem to be perfect solutions for not only hosting big data workloads but also for analyzing them.

\section{Conclusions and Future Work}\label{section.conclusion}
Cell phones can be viewed as effective sensors to help collect rich spatiotemporal data about human mobility patterns. Accessing these anonymous data enable us to study people's movement, measure the similarity of their travels, and track their mobility behaviors. In this work, we have studied the ways that mobile phone data can be treated and the existing applications and methods are reviewed. We have investigated these approaches, their relevant advantages, and drawbacks to present a taxonomy of capabilities. Predominantly, the mobility of people has been considered within mobile networks domain in order to decrease management cost. Nonetheless, in recent studies, most researchers have focused on human mobility and its impact on various social issues. They have also concentrated on users' routines and their movement habits in order to improve mobile location-based services. Typically, in such services, academic research has been focused on a single user, while human mobility research has considered human groups and their consequence mobility patterns. Perception about regularities of groups can be important in the fields of urban infrastructure planning, travel forecasting, and social relations. When it comes to the monitoring of mobile phone location data, the data representation is a relatively immature area and implemented techniques for displaying/exploring routes, velocities, directions, and volumes are rather limited. For traffic management purposes it is needed that the current monitoring system merges with the monitoring system based on cell data. The visualization of unconstrained movements within a region, as opposed to movements between pre-defined regions or along pre-defined routes, should be more explored. More research should be undertaken on the application of mobile phone data in infrastructure planning, public transportation, and disaster/rare event management \cite{Ding2018objectives,Lu2019Clustering,Lu2019Novel}. Given the exponential growth of sensors' data, it requires computational infrastructure to maintain and process large-scale datasets. A remarkable challenge is that this expansion rate of data production surpasses the ability of data processing methods. The application of big data frameworks and analyzing mobile phone data in real-time can open up ranges of opportunities to understand diverse social activities \cite{kitchin2013big}. They have the potential to improve evidence-based responses to various events (i.e., natural disasters, disease outbreaks, and emergencies) and better management of these circumstances. To meet the storage requirements and processing, Cloud \cite{Yuan2019Multi,Yuan2019Spatiotemporal,Yuan2019Temporal,Yuan2019Spatial,Li2019Performance} is a promising paradigm, capable of providing a dynamic, flexible, resilient and cost-effective infrastructure, not only to provide sufficient infrastructures for processing and storing but also for analysis purposes \cite{ghahramani2017toward,Xu2018Internet,ghahramani2017analysis,ghahramani2018Spatio-Temporal}. Moreover, the recent Fog and Edge Computing paradigms promise to provide the benefits of Cloud without incurring its problems (e.g., high latency). Future work should focus on the application of such frameworks to perform analysis to study the characteristics of mobile phone data to retrieve knowledge in an intelligent manner \cite{Casadei2019development,Liu2017Scalable,Du2019Contract,Jiang2017big,Fan2019cost,Zhang2018security,ghahramaniAI}.





\ifCLASSOPTIONcaptionsoff
  \newpage
\fi

\begin{IEEEbiography}[{\includegraphics[width=1in,height=1.25in,clip,keepaspectratio]{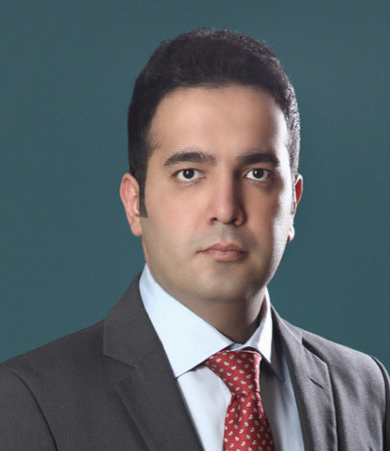}}]{Mohammadhossein Ghahramani}
obtained the B.S. degree and M.S. degree in Information Technology Engineering from Amirkabir University of Technology - Tehran Polytechnic, Iran, and Ph.D. degree in Computer Technology and Application from Macau University of Science and Technology, Macau in 2018. He was a technical manager and senior data analyst of the Information Center of Institute for Research in Fundamental Sciences from 2008 to 2014. He is currently a Post-Doctoral Research Fellow at University College Dublin (UCD), Ireland. He also is a member of the Insight Centre for Data Analytics at UCD. His research interests are focused on Machine Learning, Artificial Intelligence, Big Data, Smart Cities, and IoT. Dr. Ghahramani was a recipient of the Best Student Paper Award of 2018 IEEE International Conference on Networking, Sensing and Control. He has served as a reviewer of over ten Trans. journals including IEEE Transactions on Cybernetics, IEEE Transactions on Neural Networks and Learning Systems and IEEE Transactions on Industrial Informatics.\end{IEEEbiography}

\begin{IEEEbiography}[{\includegraphics[width=1in,height=1.25in,clip,keepaspectratio]{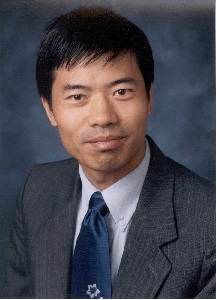}}]{MengChu Zhou}
(S'88-M'90-SM'93-F'03) received his B.S. degree in Control Engineering from Nanjing University of Science and Technology, Nanjing, China in 1983, M.S. degree in Automatic Control from Beijing Institute of Technology, Beijing, China in 1986, and Ph. D. degree in Computer and Systems Engineering from Rensselaer Polytechnic Institute, Troy, NY in 1990.  He joined New Jersey Institute of Technology (NJIT), Newark, NJ in 1990, and is now a Distinguished Professor of Electrical and Computer Engineering. His research interests are in Petri nets, intelligent automation, Internet of Things, big data, web services, and intelligent transportation.  He has over 800 publications including 12 books, over 500 journal papers (over 400 in IEEE transactions), 12 patents and 29 book-chapters. He is the founding Editor of IEEE Press Book Series on Systems Science and Engineering and Editor-in-Chief of IEEE/CAA Journal of Automatica Sinica. He is a recipient of Humboldt Research Award for US Senior Scientists from Alexander von Humboldt Foundation, Franklin V. Taylor Memorial Award and the Norbert Wiener Award from IEEE Systems, Man and Cybernetics Society. He is a life member of the Chinese Association for Science and Technology-USA and served as its President in 1999. He is a Fellow of International Federation of Automatic Control (IFAC), American Association for the Advancement of Science (AAAS) and Chinese Association of Automation (CAA).

\end{IEEEbiography}
\begin{IEEEbiography}[{\includegraphics[width=1in,height=1.25in,clip,keepaspectratio]{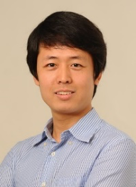}}]{Gang Wang}
received his B. S. Degree in Information Security, and M. S. Degree in Electronics and Communications Engineering from Beijing University of Posts and Telecommunications, Beijing, China, in 2007 and 2013, respectively. He is currently pursuing the Ph.D. degree in computer technology and application with Macau University of Science and Technology, Taipa, Macau. His research interests include data minning and applications, consumer behaviour and customer value in the telecom environment.
\end{IEEEbiography}

\vfill
\end{document}